\documentclass[twocolumn]{aastex63}

\usepackage{textcomp, gensymb, amsmath}
\usepackage{float}
\usepackage[T1]{fontenc} 
\usepackage[utf8]{inputenc}
\newcommand{\unit}[1]{\ensuremath{\, \mathrm{#1}}}

\shorttitle{Helium in the V1298 Tau System}
\shortauthors{Vissapragada et al.}

\begin{document}

\title{A Search for Planetary Metastable Helium Absorption in the V1298 Tau System}

\correspondingauthor{Shreyas~Vissapragada}
\email{svissapr@caltech.edu}

\suppressAffiliations

\author[0000-0003-2527-1475]{Shreyas~Vissapragada}
\affiliation{Division of Geological and Planetary Sciences, California Institute of Technology, 1200 East California Blvd, Pasadena, CA 91125, USA}

\author[0000-0001-7409-5688]{Guðmundur~Stefánsson}
\altaffiliation{Henry Norris Russell Fellow}
\affil{Department of Astrophysical Sciences, Princeton University, 4 Ivy Lane, Princeton, NJ 08540, USA}

\author[0000-0002-0371-1647]{Michael~Greklek-McKeon}
\affil{Division of Geological and Planetary Sciences, California Institute of Technology, 1200 East California Blvd, Pasadena, CA 91125, USA}

\author[0000-0002-9584-6476]{Antonija~Oklop{\v{c}}i{\'c}}
\affil{Anton Pannekoek Institute of Astronomy, University of Amsterdam, Science Park 904, 1098 XH Amsterdam, Netherlands}

\author[0000-0002-5375-4725]{Heather~A.~Knutson}
\affil{Division of Geological and Planetary Sciences, California Institute of Technology, 1200 East California Blvd, Pasadena, CA 91125, USA}

\author[0000-0001-8720-5612]{Joe P.\ Ninan}
\affil{Department of Astronomy \& Astrophysics, The Pennsylvania State University, 525 Davey Laboratory, University Park, PA, 16802, USA}
\affil{Center for Exoplanets and Habitable Worlds, 525 Davey Laboratory, University Park, PA, 16802, USA}

\author[0000-0001-9596-7983]{Suvrath Mahadevan}
\affil{Department of Astronomy \& Astrophysics, The Pennsylvania State University, 525 Davey Laboratory, University Park, PA, 16802, USA}
\affil{Center for Exoplanets and Habitable Worlds, 525 Davey Laboratory, University Park, PA, 16802, USA}

\author[0000-0003-4835-0619]{Caleb I. Ca\~nas}
\altaffiliation{NASA Earth and Space Science Fellow}
\affil{Department of Astronomy \& Astrophysics, The Pennsylvania State University, 525 Davey Laboratory, University Park, PA, 16802, USA}
\affil{Center for Exoplanets and Habitable Worlds, 525 Davey Laboratory, University Park, PA, 16802, USA}

\author[0000-0003-1728-8269]{Yayaati~Chachan}
\affil{Division of Geological and Planetary Sciences, California Institute of Technology, 1200 East California Blvd, Pasadena, CA 91125, USA}

\author[0000-0001-9662-3496]{William D. Cochran}
\affil{Center for Planetary Systems Habitability, The University of Texas at Austin, 2515 Speedway, Austin, TX 78712, USA}
\affil{McDonald Observatory and Department of Astronomy, The University of Texas at Austin, 2515 Speedway, Austin, TX 78712, USA}

\author[0000-0001-6588-9574]{Karen~A.~Collins}
\affil{Center for Astrophysics $|$ Harvard \& Smithsonian, 60 Garden Street, Cambridge, MA 02138, USA}

\author[0000-0002-8958-0683]{Fei~Dai}
\affil{Division of Geological and Planetary Sciences, California Institute of Technology, 1200 East California Blvd, Pasadena, CA 91125, USA}

\author[0000-0001-6534-6246]{Trevor~J.~David}
\affil{Center for Computational Astrophysics, Flatiron Institute, New York, NY 10010, USA}
\affil{Department of Astrophysics, American Museum of Natural History, New York, NY 10024, USA}

\author[0000-0003-1312-9391]{Samuel Halverson}
\affil{Jet Propulsion Laboratory, 4800 Oak Grove Drive, Pasadena, CA 91109, USA}

\author[0000-0002-6629-4182]{Suzanne L. Hawley}
\affil{University of Washington Department of Astronomy, 3910 15th Ave NE, Seattle, WA 98195, USA}

\author[0000-0003-1263-8637]{Leslie Hebb}
\affiliation{Department of Physics, Hobart and William Smith Colleges, 300 Pulteney Street, Geneva, NY, 14456, USA}

\author[0000-0001-8401-4300]{Shubham Kanodia}
\affil{Department of Astronomy \& Astrophysics, The Pennsylvania State University, 525 Davey Laboratory, University Park, PA, 16802, USA}
\affil{Center for Exoplanets and Habitable Worlds, 525 Davey Laboratory, University Park, PA, 16802, USA}

\author[0000-0001-7458-1176]{Adam F. Kowalski}
\affil{Department of Astrophysical and Planetary Sciences, University of Colorado Boulder, 2000 Colorado Ave,
Boulder, CO 80305, USA}
\affil{National Solar Observatory, University of Colorado Boulder, 3665 Discovery Drive, Boulder, CO 80303, USA}
\affil{Laboratory for Atmospheric and Space Physics, University of Colorado Boulder, 3665 Discovery Drive, Boulder, CO 80303, USA}

\author[0000-0002-4881-3620]{John~H.~Livingston}
\affiliation{Department of Astronomy, University of Tokyo, 7-3-1 Hongo, Bunkyo-ku, Tokyo 113-0033, Japan}

\author[0000-0001-8222-9586]{Marissa Maney}
\affil{Department of Astronomy \& Astrophysics, The Pennsylvania State University, 525 Davey Laboratory, University Park, PA, 16802, USA}

\author[0000-0001-5000-1018]{Andrew J. Metcalf}
\affiliation{Space Vehicles Directorate, Air Force Research Laboratory, 3550 Aberdeen Ave. SE, Kirtland AFB, NM 87117, USA}
\affiliation{Time and Frequency Division, National Institute of Standards and Technology, 325 Broadway, Boulder, CO 80305, USA} 
\affiliation{Department of Physics, University of Colorado, 2000 Colorado Avenue, Boulder, CO 80309, USA}

\author[0000-0002-4404-0456]{Caroline Morley}
\affil{McDonald Observatory and Department of Astronomy, The University of Texas at Austin, 2515 Speedway, Austin, TX 78712, USA}
\affil{Center for Planetary Systems Habitability, The University of Texas at Austin, 2515 Speedway, Austin, TX 78712, USA}

\author[0000-0002-4289-7958]{Lawrence W. Ramsey}
\affil{Department of Astronomy \& Astrophysics, The Pennsylvania State University, 525 Davey Laboratory, University Park, PA, 16802, USA}
\affil{Center for Exoplanets and Habitable Worlds, 525 Davey Laboratory, University Park, PA, 16802, USA}

\author[0000-0003-0149-9678]{Paul Robertson}
\affil{Department of Physics and Astronomy, The University of California, Irvine, Irvine, CA 92697, USA}

\author[0000-0001-8127-5775]{Arpita Roy}
\affil{Space Telescope Science Institute, 3700 San Martin Drive, Baltimore, MD 21218, USA}
\affil{Department of Physics and Astronomy, Johns Hopkins University, 3400 N Charles St, Baltimore, MD 21218, USA}

\author[0000-0002-5547-3775]{Jessica~Spake}
\affil{Division of Geological and Planetary Sciences, California Institute of Technology, 1200 East California Blvd, Pasadena, CA 91125, USA}

\author[0000-0002-4046-987X]{Christian Schwab}
\affil{Department of Physics and Astronomy, Macquarie University, Balaclava Road, North Ryde, NSW 2109, Australia}

\author[0000-0002-4788-8858]{Ryan C. Terrien}
\affil{Department of Physics and Astronomy, Carleton College, One North College Street, Northfield, MN 55057, USA}

\author[0000-0002-1481-4676]{Samaporn~Tinyanont}
\affiliation{Department of Astronomy and Astrophysics, University of California, Santa Cruz, CA 95064, USA}

\author[0000-0002-1871-6264]{Gautam~Vasisht}
\affil{Jet Propulsion Laboratory, California Institute of Technology, 4800 Oak Grove Dr, Pasadena, CA 91109, USA}

\author[0000-0001-9209-1808]{John Wisniewski}
\affiliation{Homer L. Dodge Department of Physics and Astronomy, University of Oklahoma, 440 W. Brooks Street, Norman, OK 73019, USA}

\begin{abstract}
Early in their lives, planets endure extreme amounts of ionizing radiation from their host stars. For planets with primordial hydrogen and helium-rich envelopes, this can lead to substantial mass loss. Direct observations of atmospheric escape in young planetary systems can help elucidate this critical stage of planetary evolution. In this work, we search for metastable helium absorption---a tracer of tenuous gas in escaping atmospheres---during transits of three planets orbiting the young solar analogue V1298 Tau. We characterize the stellar helium line using HET/HPF, and find that it evolves substantially on timescales of days to months. The line is stable on hour-long timescales except for one set of spectra taken during the decay phase of a stellar flare, where absoprtion increased with time. Utilizing a beam-shaping diffuser and a narrowband filter centered on the helium feature, we observe four transits with Palomar/WIRC: two partial transits of planet d ($P = 12.4$ days), one partial transit of planet b ($P = 24.1$ days), and one full transit of planet c ($P = 8.2$ days). We do not detect the transit of planet c, and we find no evidence of excess absorption for planet b, with $\Delta R_\mathrm{b}/R_\star<0.019$ in our bandpass. We find a tentative absorption signal for planet d with $\Delta R_\mathrm{d}/R_\star = 0.0205\pm0.054$, but the best-fit model requires a substantial (-100$\pm$14 min) transit-timing offset on a two-month timescale. Nevertheless, our data suggest that V1298 Tau d may have a high present-day mass-loss rate, making it a priority target for follow-up observations.
\end{abstract}


\section{Introduction} \label{sec:intro}
The atmospheres of close-in exoplanets evolve substantially over their lifetimes. As planets cool and contract after formation, their extended atmospheres are subject to intense high-energy radiation from their young, active host stars on timescales of up to a Gyr \citep{Owen19, King21}. This radiation heats the planetary thermosphere via photoionization, and can launch a hydrodynamic wind that carries mass away from the planet \citep{MurrayClay09}. Additionally, the planet's own cooling interior may itself power mass loss as the envelope opacity decreases \citep{Ginzburg18, Gupta19}. Regardless of the mechanism, the impact of early atmospheric mass loss can be observed in population-level studies of older planets.  Over the last decade, this phenomenon has been invoked to explain both the radius valley \citep{Lopez13, Owen13, Fulton17, Owen17, Fulton18, VanEylen18, HardegreeUllman20} and the lower boundary of the hot Neptune desert \citep{Lundkvist16, Mazeh16, Owen18}.

In order to accurately interpret population-level trends, we must rely on models to predict the cumulative atmospheric mass loss from individual planets.  Observations of present-day mass loss rates for close-in planets provide an invaluable test of these models.  For nearby systems ($\lesssim100$~pc) that are not fully obscured by absorption from the interstellar medium, transmission spectroscopy in the wings of the Lyman-$\alpha$ line can reveal escaping high-velocity neutral hydrogen \citep[e.g.][]{VidalMadjar03, Lecavelier10, Ehrenreich15, Bourrier18}. Near-infrared observations of metastable helium absorption at 1083~nm provide a complementary probe of atmospheric escape \citep{Oklopcic18, Spake18}. The accessibility of the metastable helium triplet from the ground makes it well-suited to surveys using high-resolution spectroscopy \citep{Allart18, Nortmann18, Salz18, Allart19, AlonsoFloriano19, Kirk20, Ninan20, Palle20} as well as narrowband photometry \citep{Vissapragada20, Paragas21}.  Helium absorption has also been detected in two systems using high-precision $R\sim100$ spectroscopy with the \textit{Hubble Space Telescope} \citep{Mansfield18, Spake18}, although the sensitivity of these measurements is limited by their relatively low spectral resolution.  

Over the past few years, these observations have significantly expanded the sample of transiting planets with well-constrained present-day mass loss rates.  However, all of the systems observed to date have estimated ages of a few Gyr. By that time, most of the major processes that sculpt these systems---including atmospheric erosion---have largely concluded. It would be preferable to observe atmospheric escape in young planetary systems, as this would allow us to test our understanding of mass loss physics during the epoch when most atmopsheric mass loss is predicted to occur. There are currently only seven confirmed young ($<100$~Myr) transiting planet systems: K2-33 \citep{David16, Mann16}, V1298 Tau \citep{David19a, David19b}, DS Tuc A \citep{Newton19}, TOI 837 \citep{Bouma20}, AU Mic \citep{Plavchan20}, TOI 942 \citep{Carleo20, Zhou21}, and HIP 67522 \citep{Rizzuto20}. The expected magnitude of the outflows from these young planets depends on their gravitational potentials \citep[e.g.][]{Hirano20}, which can be calculated using their measured masses and radii.  Unfortunately, it is difficult to obtain precise radial velocities (RVs) for young stars \citep[e.g.][]{Beichman19, Plavchan20, Klein20}, which severely limits our knowledge of the planetary masses.

Of the aforementioned young transiting systems, V1298 Tau is a uniquely favorable target for observations of atmospheric escape. First, it is one of only two K stars in the sample \citep{David19a}. K-type stars are optimal for observations of metastable helium, as they have a favorable ratio of EUV to mid-UV flux, which sets the level population in the metastable state \citep{Oklopcic19}. Although early M stars have a similarly favorable flux ratio, K stars output a larger total EUV flux than M stars, resulting in more helium ionization and subsequent recombination into the metastable state. We note that because V1298 Tau is a pre-main sequence star, the radiative physics may differ somewhat from the more mature K dwarfs considered in \cite{Oklopcic19}. The V1298 Tau system is also dynamically unique; although the orbital period of the outermost planet is poorly constrained, the orbital periods of the three interior planets are close to a 2:3:6 chain of mean-motion resonances, with planet c orbiting V1298 Tau every 8.2 days, planet d every 12.4 days, and planet b every 24.1 days \citep{David19b}. This should allow for strong constraints on the planetary masses using the transit-timing variation (TTV; Livingston et al. in prep.) technique, circumventing the difficulties of RV mass measurements for this young system.

In this work, we use the metastable helium triplet at 1083~nm to search for atmospheric outflows from the three innermost planets in the V1298 Tau system. In Section~\ref{sec:obs}, we describe our narrowband helium transit observations with the Wide-field InfraRed Camera \citep[WIRC;][]{Wilson03} on the Hale 200" telescope at Palomar Observatory, as well as complementary observations obtained with the Habitable-zone Planet Finder \citep[HPF;][]{mahadevan2012,mahadevan2014} near-infrared spectrograph on the 10m Hobby-Eberly Telescope (HET). In Section~\ref{sec:model1} we fit the resulting transit light curves, tentatively detecting an increased radius ratio in the helium bandpass for planet d. We discuss our results in Section~\ref{sec:disc}, and we conclude in Section~\ref{sec:conc} by summarizing the implications of our results and detailing the highest-priority future observations needed for confirmation.

% -----------------------------------------------------
% -----------------------------------------------------
% -----------------------------------------------------
\section{Observations} \label{sec:obs}

% -----------------------------------------------------
\subsection{Palomar/WIRC}
We observed two partial transits of V1298 Tau d, one full transit of V1298 Tau c, and one partial transit of V1298 Tau b between 2020 October and 2021 January. A summary of our observations is given in Table~\ref{table:log}. We utilized a beam-shaping diffuser \citep{Stefansson17, Vissapragada20a} and a narrowband helium filter for these observations with an exposure time of 30~s. The diffuser molds the stellar PSFs into a top-hat shape with a FWHM of 3\arcsec, mitigating noise stemming from time-correlated variations in the stellar point spread function and improving our overall observing efficiency by allowing for longer integration times. Our custom narrowband helium filter is centered at 1083.3~nm with a FWHM of 0.635~nm. In \cite{Vissapragada20}, we presented commissioning observations using this filter where we reproduced the helium absorption signal reported by \citet{Nortmann18} for WASP-69b. 

\begin{deluxetable*}{ccccccccc}[t!]
\tabletypesize{\scriptsize}
\tablecaption{Summary of V1298 Tau observations analyzed in this work \label{table:log}}
\tablehead{\colhead{Instrument} & \colhead{Planet} & \colhead{Coverage} & \colhead{Date}        & \colhead{Start Time} & \colhead{End Time} & \colhead{Exposures} & \colhead{Exposure Time} & \colhead{Start/Min/End Airmass}}
\startdata
WIRC                            & d                & Ingress            & 2020 Oct 8            & 06:09:49             & 13:17:25           & 669                 & 30~s                    & 2.10/1.03/1.24 \\
WIRC                            & d                & Egress             & 2020 Nov 27           & 03:46:32             & 11:48:57           & 759                 & 30~s                    & 1.57/1.03/1.93 \\
WIRC                            & c                & Full               & 2020 Dec 17           & 02:39:14             & 11:00:35           & 784                 & 30~s                    & 1.50/1.03/2.38 \\
WIRC                            & b                & Egress             & 2021 Jan 1            & 02:40:03             & 09:58:12           & 660                 & 30~s                    & 1.23/1.03/2.32 \\
HPF                             & c                & In-transit         & 2020 Oct 12           & 06:38:12             & 07:42:41           & 6                   & 618~s                   & 1.33/1.13/1.13 \\
HPF                             & --               & Out-of-transit     & 2020 Oct 13           & 11:33:20             & 12:05:18           & 3                   & 618~s                   & 1.14/1.14/1.23 \\
HPF                             & --               & Out-of-transit     & 2020 Oct 14           & 11:42:03             & 12:14:01           & 3                   & 618~s                   & 1.18/1.28/1.18 \\
HPF                             & --               & Out-of-transit     & 2020 Nov 14           & 09:33:42             & 10:32:51           & 6                   & 565~s                   & 1.16/1.16/1.37 \\
\enddata
\tablecomments{All dates and times are UT.}
\end{deluxetable*}

As described in \cite{Vissapragada20}, the filter bandpass shifts with angle of incidence (AOI).  This means that only a small fraction of the $8.7\arcmin\times8.7\arcmin$ field of view of our camera has an effective bandpass centered on the the metastable helium feature. We calibrated this effect at the start of each night by illuminating the detector with light from a helium arc lamp (which emits in the 1083~nm line) and placing V1298 Tau on the resultant bright semicircular region where the effective filter bandpass is aligned with the metastable helium feature. Before beginning the exposure sequence on each night, we also performed a seven-point dither to construct a sky background frame.

Conditions were clear for the first night of observations on 2020 October 8 UT, but on 2020 November 27 UT, the seeing was exceptionally poor for Palomar (sometimes exceeding 3\arcsec). On the night of 2020 December 17 UT, there were rapid transparency variations due to thin cirrus coverage for most of the observation. On the night of 2021 January 1 UT, conditions were mostly clear, with some thin cirrus coverage appearing in the final hour. Due to a software crash we lost about 15 minutes of data in the middle of this final night.

The data were dark-subtracted, flat-fielded, and corrected for bad pixels using the method detailed in \citet{Tinyanont19}. As in \citet{Vissapragada20}, we also corrected the non-uniform background (which arises from telluric OH emission lines) by median-scaling the dithered background frame in 10~px radial steps beginning from the filter zero point at the top of the detector, where light has encountered the filter at normal incidence. After image calibration, we performed aperture photometry on the target star and two comparison stars of similar brightness (HD 284153 and HD 284154) using the \texttt{photutils} package \citep{Bradley19}, trying apertures with radii ranging from 5 to 20 px in 1 px steps (the pixel scale for WIRC is 0\farcs 25/px). We allowed the aperture centers to shift in each image in order to track variations in the telescope pointing, and found average pointing shifts of 2-3~px on all nights. We subtracted any residual local background using an annulus around each source with an inner radius of 25 px and an outer radius of 50 px. 

We pre-processed the light curves for all aperture sizes by clipping 5$\sigma$ outliers using a moving median filter with a window size of 11. We then selected the optimal aperture size for each night of data by minimizing the per-point rms of the median-filtered photometry. We preferred an 11~px aperture for the 2020 October 8 night, a 15~px aperture for the 2020 November 27 night, a 7~px aperture for the 2020 December 17 night, and a 10~px aperture for the 2021 January 1 night. 

\subsection{HPF/HET}
We obtained high resolution spectra of V1298 Tau with the Habitable-zone Planet Finder (HPF) spectrograph \citep[][]{mahadevan2012,mahadevan2014}, a high-resolution ($R\sim55,000$) temperature-stabilized \citep{stefansson2016} fiber-fed \citep{kanodia2018} spectrograph on the 10m Hobby Eberly Telescope at McDonald Observatory. HPF operates in the near-infrared (NIR) covering the $z$, $Y$, and $J$ bands from 810-1280~nm, and fully resolves the He 1083~nm line. We observed V1298 Tau with HPF on 2020 October 12-14 UT and 2020 November 14 UT. The October 12 observation coincided with a transit of V1298 Tau c. For the October observations, we used an exposure time of 617.7~s, and for the November observations we used an exposure time of 564.5~s. We collected 6 exposures on the first night, 3 exposures on the second night, 3 exposures on the third night, and 6 exposures on the fourth night, with a median S/N of 143 per 1D extracted pixel at $1\unit{\mu m}$.

The HPF 1D spectra were processed using the procedures described in \cite{ninan2018}, \cite{kaplan2019}, \cite{metcalf2019}, and \cite{stefansson2020}. Following \cite{Ninan20}, we elected to observe V1298 Tau without the simultaneous HPF Laser Frequency Comb (LFC) calibrator, to minimize any possible impact from scattered light in the target star fiber. We deblazed the spectra using a combination of a static flat HPF exposure and a simultaneous low-order polynomial fit to account for residual low-order spectral differences for the different HPF visits. To correct for OH-sky emission, we estimated the sky background using the simultaneous HPF sky fiber and subtracted this background from the target star spectrum following \cite{Ninan20}. 

We correct for telluric absorption by using \texttt{molecfit} \citep{smette2015,kausch2015} to fit telluric features in the deblazed and continuum normalized HPF spectra. It is important to obtain an accurate correction for the time-varying telluric absorption, as there are two telluric water lines that partially overlap with the He 1083~nm feature. For the \texttt{molecfit} fit, we used a Gaussian kernel to describe the HPF PSF. Although in actuality the HPF PSF shape is more complex, we tried modeling the PSF as a combination of a Gaussian and a Lorentzian and found that this did not improve the quality of the telluric correction.  We therefore opted to use the simpler Gaussian fit in our final analysis. We restricted the fit to water lines, and allowed for the time-varying shape of these telluric features by fitting each spectrum independently. We also inflated the uncertainties in the vicinity of the telluric region by a factor of 1.2 to match the uncertainty estimate with the as-observed additional scatter due to imperfections in the telluric correction.

\begin{figure*}[ht!]
\centering
\includegraphics[width=1.0\textwidth]{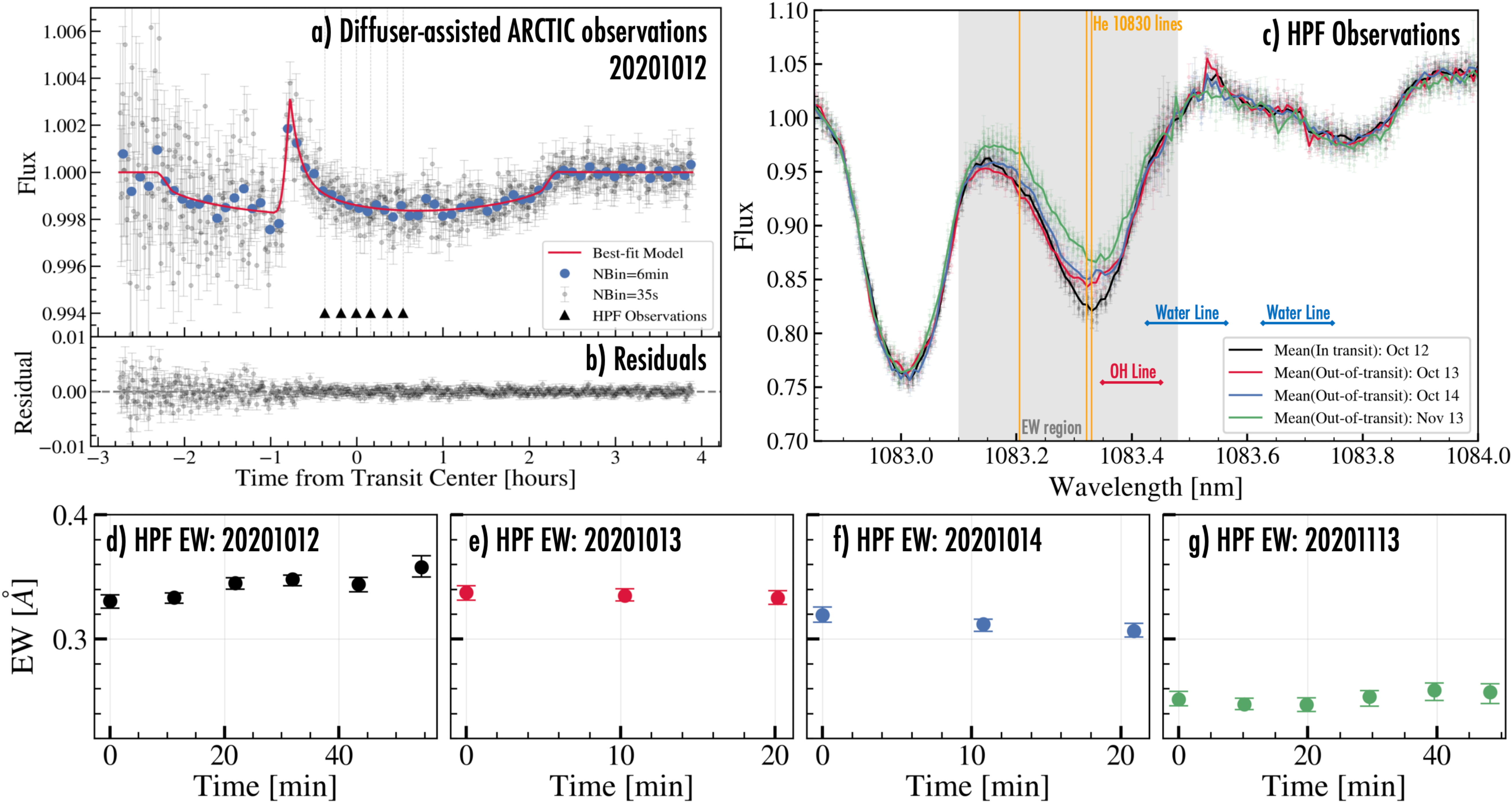}
\caption{a) Diffuser-assisted transit observations of V1298 Tau c. Unbinned data are shown in black, and 6~min binned data are shown in blue. A large stellar flare is visible during the transit. The best-fit transit plus flare model is shown in red. The model and data are shown after removing the best-fit Gaussian Process correlated noise model. The black triangles denote the timing of the HPF observations obtained during the transit. b) Residuals from transit fit. c) Spectroscopic observations obtained with HPF of the He 1083~nm triplet (orange lines show the He 1083~nm triplet in the stellar rest frame). The averaged spectrum in black was obtained during the transit shown in the left panel, while averaged spectra from HPF visits on other nights are shown as colored points and lines. The locations of the OH and water lines are indicated in red and blue, respectively. d-g) Equivalent width (EW) measurements of the He 1083~nm line inside the grey region shown in panel c. We see a linear increase in the EWs during the transit (black points in d). The EWs remain stable over $\sim$10~min timescales in the other visits, but change on timescales of days to months.}
\label{spectra}
\end{figure*}

Figure \ref{spectra} shows the resulting spectra after the telluric correction is applied. After applying the telluric correction, we calculated the equivalent width of the line in a region from 1083.1--1083.48~nm (see grey region in Figure \ref{spectra}b), which spans the full extent of the observed variation in the He 1083~nm line. Also shown is a light curve of V1298 Tau c taken on 2020 October 12 from the ARCTIC \citep[Astrophysical Research Consortium Telescope Imaging Camera;][]{huehnerhoff2016} imager on the Astrophysical Research Consortium (ARC) 3.5m Telescope at Apache Point Observatory. The light curve exhibits a clear stellar flare 19~min before the HPF spectroscopic observations started. These observations and their associated modeling are described in detail in Livingston et al. (in prep.) Briefly, data were taken with a beam-shaping diffuser and narrow-band ($30 \unit{nm}$) filter with minimal telluric absorption lines centered around $857 \unit{nm}$ \citep[see e.g.,][]{Stefansson17,stefansson2018}. The flare, transit, and correlated noise were modeled using the \texttt{allesfitter} package \citep{gunther2019, gunther2021}, using the \citet{davenport2014} parameterization for the flare and a Matern-3/2 kernel from \texttt{celerite} \citep{foremanmackey2017} to describe the correlated noise.

% -----------------------------------------------------
% -----------------------------------------------------
% -----------------------------------------------------
\section{Light-curve Modeling} \label{sec:model1}

% -----------------------------------------------------
\subsection{Palomar/WIRC}
We fit the aperture-optimized WIRC photometry using the \texttt{exoplanet} package \citep{ForemanMackey20}. The priors for our model parameters are listed in Table~\ref{table1}. For each WIRC dataset, we modeled the light curve with a limb-darkened transit model (normalized to one) from \texttt{starry} \citep{Luger19} multiplied by a systematics model and then added to a baseline model. We fixed the mean ephemerides to the best-fit solution from Livingston et al. (in prep.), and used \texttt{TTVOrbit} to allow for an offset in the transit timing for each data set. For planet c, we placed a restrictive prior $\mathcal{N}(0~\mathrm{min}, 30~\mathrm{min}$) on the offset $\Delta T_\mathrm{c}$ to avoid fitting correlated features at the edges of the light curve. For planet b, we used a wide uniform prior $\mathcal{U}(-120~\mathrm{min}, 120~\mathrm{min}$) on the offset $\Delta T_\mathrm{b}$. The transit timing for planet d was measured independently by the Las Cumbres Observatory Global Telecope \citep[LCOGT][]{Brown13} 1~m Sinistro imager on 2020 October 8 UT (Livingston et al. in prep.), so we fit that transit time $t_\mathrm{ingress}$ with a normal prior centered on the LCOGT transit time, and allowed for a TTV offset between the ingress and egress epochs with a uniform prior of $\mathcal{U}(-120~\mathrm{min}, 120~\mathrm{min})$. 

We placed a uniform prior on the impact parameter $b$, which we sample using the algorithm from \cite{Espinoza18}.  We fit this parameter jointly with detrended light curves from \textit{K2} \citep{David19b} in order to ensure that the final light curves have the correct transit duration, as we could only achieve partial phase coverage in most of our observations. We fit the the radius ratio in the WIRC bandpass with a wide uniform prior $\mathcal{U}(0, 0.3)$ for each planet.  We use the difference between the WIRC radius ratio and the radius ratio in the $K2$ bandpass \citep{David19b} as a measure of the excess absorption in the helium feature due to the presence of an extended atmosphere, which we label $\Delta R_\mathrm{p}/R_\star$. We additionally included normal priors on the stellar mass and radius from \citet{David19b}, and fit for the quadratic limb-darkening coefficients $(u_1, u_2)$ using the \citet{Kipping13} prescription. Finally, we fit a photometric jitter term $\log(\sigma_\mathrm{extra})$ for each dataset to quantify the average scatter in excess of the photon noise.

For the baseline, we initially used a linear function $a_1x' + a_0$ for each light curve, where $a_i$ are free parameters with uniform priors $\mathcal{U}(-1, 1)$ for each dataset and $x' = x - \textrm{med}(x)$ are the median-normalized BJD observation times. For planet c, where the observing conditions were relatively poor, we found that the results exhibited significant correlated noise, so we modeled the correlated component using an additional Gaussian Process (GP) term. We used a Matern-3/2 kernel as implemented in \texttt{celerite2} \citep{ForemanMackey18}, with free parameters describing the timescale and amplitude, $\rho$ and $\sigma$.  We placed wide uniform priors of $\mathcal{U}(0, 0.3~\mathrm{days})$ and $\mathcal{U}(0, 0.1)$, respectively, on these two parameters.

For our systematic noise model we used a linear combination of detrending vectors, with each vector multiplied by a weight $w_i$ and summed to generate the systematics model. We allowed each of these weights to vary as a free parameter in the fit with a uniform prior $\mathcal{U}(-1, 1)$.  There were three detrending vectors for each night, including photometry for the two comparison stars and a proxy for the time-variable telluric water absorption. As discussed in the previous section, there are two telluric water lines that overlap with the metastable helium feature.  In our photometric observations, we cannot fit for the telluric water spectrum and remove it from each individual image frame.  Uncorrected time-varying absorption in these water lines can add correlated noise to the target star photometry and bias our estimate of the transit depth. We corrected this time-varying absorption using the method detailed in \citet{Paragas21}. In short, we used the integrated OH emission line intensities in our images to track the stability of the water feature using the ratio of water-contaminated to water-uncontaminated OH emission line intensities on the detector. 

\begin{figure*}[ht!]
\centering
\includegraphics[width=\textwidth]{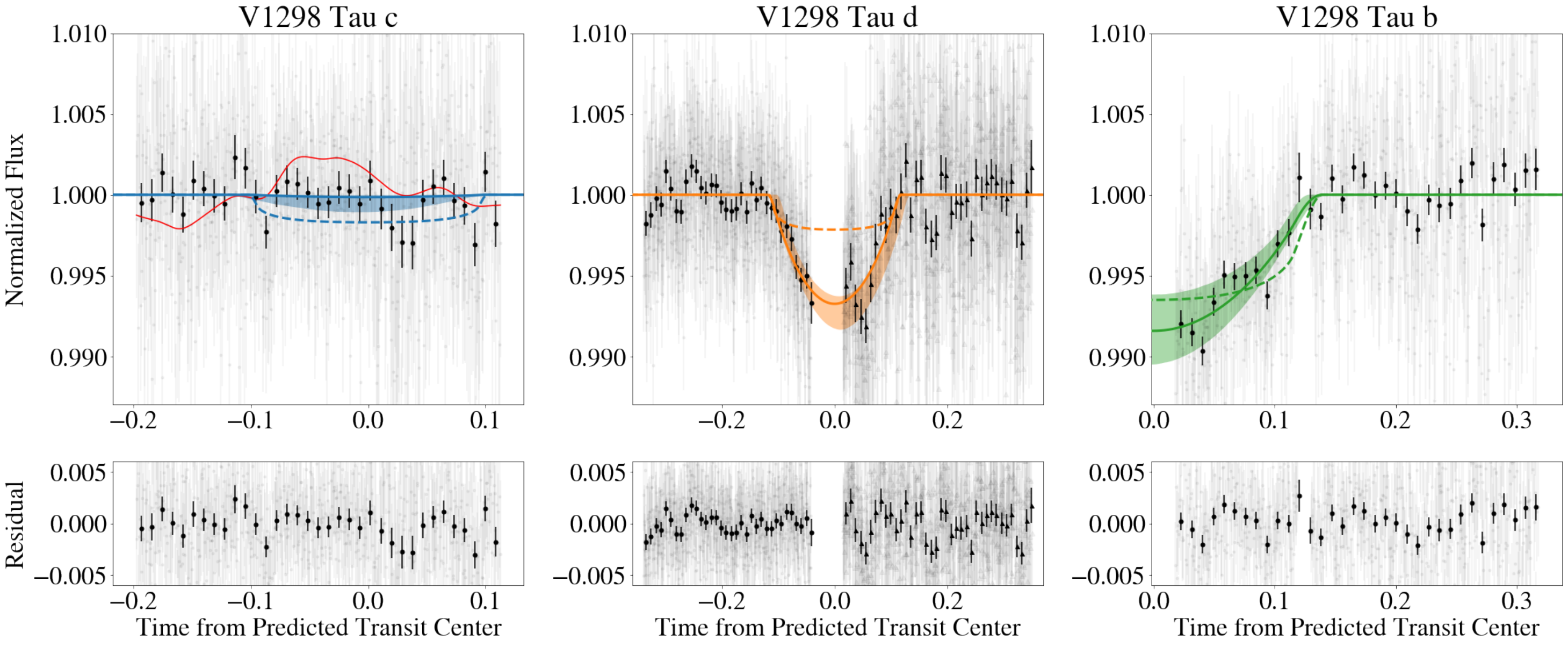}
\caption{Detrended Palomar/WIRC helium light curves and residuals for V1298 Tau b, c, and d. For planet c (top left panel), we overplot the best-fit Gaussian Process model, which was subtracted from the light curve, as a solid red line. In all panels, unbinned data are shown in gray and data binned to 10~min cadence are shown in black. The circles and triangles for V1298 Tau d denote data from the first and second nights, respectively. In the top panels, the dashed colored line represents the best-fit \textit{K2} optical light-curve model, the solid colored line indicates the best-fit WIRC light-curve model, and the shaded region represents the 68\% confidence interval on the WIRC model.} 
\label{figure1}
\end{figure*}

We used the No U-Turn Sampler \citep[NUTS; ][]{Hoffman11} implemented in \texttt{PyMC3} \citep{Salvatier16} to sample the posterior distributions for our model parameters. We ran four chains, tuning each for 1,000 steps before taking 1,500 draws from the posterior (for a total of 6,000 draws). The Gelman-Rubin statistic \citep{Gelman92} was less than 1.01 for all sampled parameters, indicating good convergence. We list the priors and posteriors for the fits in Table~\ref{table1}, and plot the phased transit data along with the best-fit models and residuals in Figure~\ref{figure1}.

% -----------------------------------------------------

\begin{deluxetable}{rlll}
\tablecaption{Priors and posteriors for V1298 Tau fits for WIRC.}
\label{table1}
\tablecolumns{4}
\tablehead{ \colhead{Parameter} & \colhead{Prior} & \colhead{Posterior} & \colhead{Units}}
\startdata
$\Delta T_\mathrm{0,c}$ & $\mathcal{N}(0, 30)$ &  $-1.3_{-29.2}^{+34.8}$ & min \\
$b_\mathrm{c}$ & \citet{Espinoza18} &  $0.14_{-0.09}^{+0.12}$ & -- \\
$R_\mathrm{c}/R_\star$ & $\mathcal{U}(0, 0.3)$ & $0.0128_{-0.0089}^{+0.0137}$ & --\\
$\Delta R_\mathrm{c}/R_\star$ & [derived] & $-0.025_{-0.009}^{+0.014}$ & -- \\
\hline
$\mathrm{t}_\mathrm{ingress}$ & $\mathcal{N}(2459131.0943, 0.0077)$ & $2459131.1088_{-0.0068}^{+0.0077}$ & BJD$_\mathrm{TDB}$  \\
$\mathrm{TTV}_\mathrm{d}$ & $\mathcal{U}(-120, 120)$ & $-101_{-12}^{+14}$ & min \\
$b_\mathrm{d}$ & \citet{Espinoza18} &  $0.124_{-0.082}^{+0.124}$ & -- \\
$R_\mathrm{d}/R_\star$ & $\mathcal{U}(0, 0.3)$ & $0.0642_{-0.0048}^{+0.0047}$ & --\\
$\Delta R_\mathrm{d}/R_\star$ & [derived] & $0.0205_{-0.0053}^{+0.0055}$ & -- \\
\hline
$\Delta T_\mathrm{0,b}$ & $\mathcal{U}(-120, 120)$ & $-56.5_{-9.9}^{+14.4}$ & min \\
$b_\mathrm{b}$ & \citet{Espinoza18} & $0.439_{-0.016}^{+0.041}$ & -- \\
$R_\mathrm{b}/R_\star$ & $\mathcal{U}(0, 0.3)$ & $0.0128_{-0.0089}^{+0.0137}$ & --\\
$\Delta R_\mathrm{b}/R_\star$ & [derived] & $0.0036_{-0.0107}^{+0.0095}$ & -- \\
\hline
$M_\star$ & $\mathcal{N}(1.101, 0.050)$ &  $1.104_{-0.045}^{+0.046}$ & $M_\Sun$\\
$R_\star$ & $\mathcal{N}(1.345, 0.056)$ & $1.333_{-0.024}^{+0.030}$ & $R_\Sun$ \\
$u_1$ & \citet{Kipping13} & $1.80_{-0.18}^{+0.11}$ & --  \\
$u_2$ & \citet{Kipping13} & $-0.851_{-0.088}^{+0.160}$ & -- \\
\enddata
\tablecomments{For brevity, we excluded the detrending vector weights, baseline coefficients, and jitter parameters from this table.}
\end{deluxetable}

% -------------------------------------------------
% -------------------------------------------------
% -------------------------------------------------
\section{Discussion}\label{sec:disc}

% -------------------------------------------------
\subsection{The Stellar Helium Line}
\label{stellar}
The spectra in Figure~\ref{spectra} show a broad, deep helium feature for V1298 Tau, reaching equivalent widths of 0.35 \AA.  The line's equivalent width varies substantially on month-long timescales, decreasing to EW = 0.2 \AA\ just one month after the first observation. We conclude that, for this 23~Myr-old pre-main sequence star, it is critical to acquire a reliable baseline measurement immediately before and after the transit.  This stands in contrast to previous literature studies of older stars, which have successfully utilized baseline measurements collected across multiple epochs to measure the strength of the planetary helium absorption during the transit \citep{Ninan20}. The overall stellar line shape appears to remain consistent across all of our observations.

Our helium transit light curves also require very strong limb-darkening in order to obtain a good fit. The Solar disk is known to exhibit strong limb-darkening in the helium line \citep[e.g.][]{Harvey77}, and this effect may be stronger for young, active stars. Additionally, there is a potential degeneracy between the model limb-darkening and the outflow geometry.  Even for modest outflows, deviations from spherical symmetry in the extended metastable helium distribution can change the morphology of the light curve \citep{Wang21a, Wang21b}. Both of these factors may contribute to the ``V-shaped'' appearance of the light curves, especially for planet d.

% ------------------------------------------------------------------
\subsection{A Stellar Flare Observed in Helium}
\begin{figure}[b!]
\centering
\includegraphics[width=0.45\textwidth]{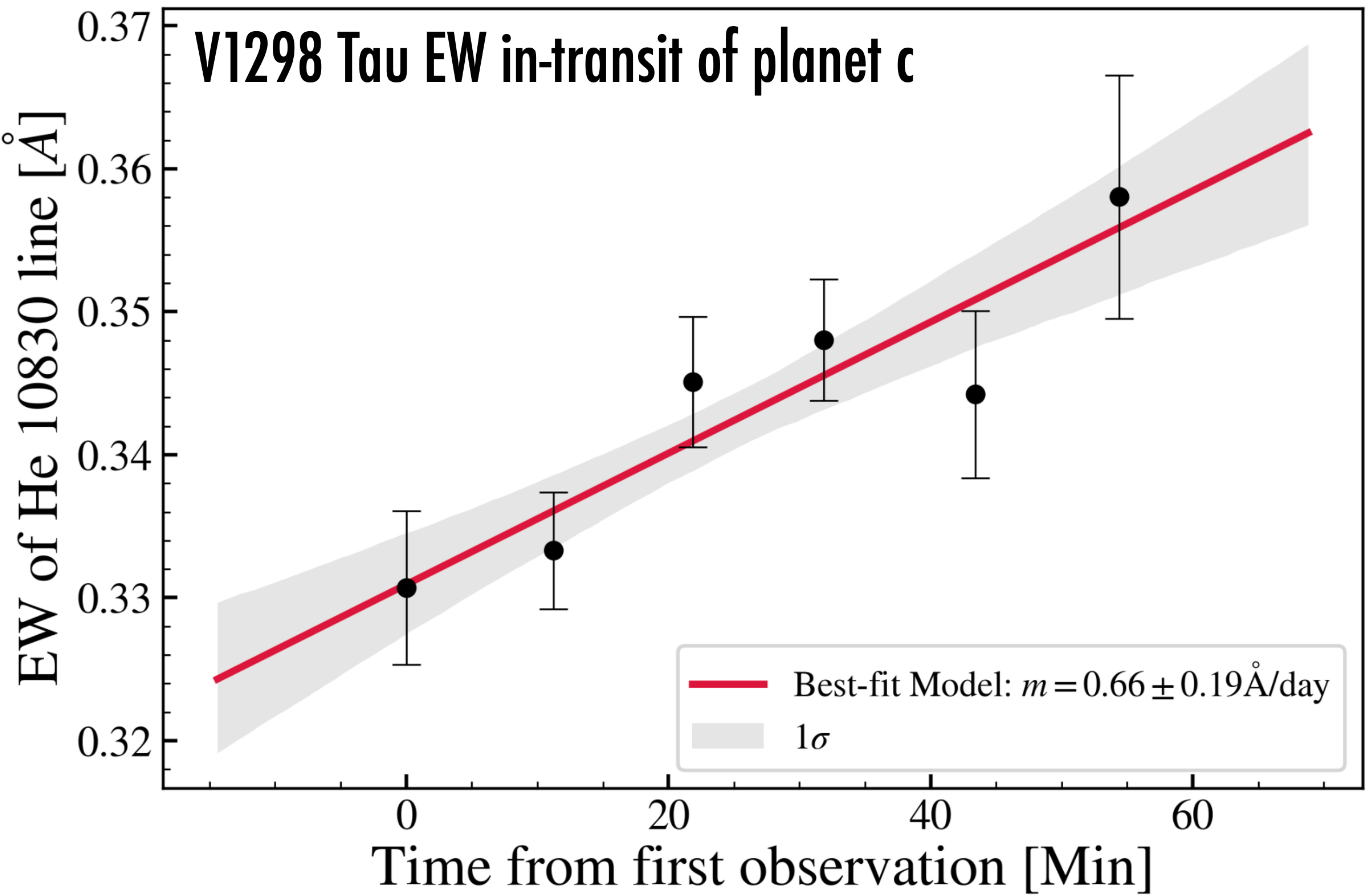}
\caption{Linear fit to the metastable helium equivalent widths in the decay phase of the flare observed on 2020 October 12. The median model is plotted as a red line, and the grey shaded region represent the 1$\sigma$ credible intervals.}
\label{increase}
\end{figure}

Six HPF spectra were obtained during a white-light flare that overlapped with a transit of V1298 Tau c. As shown in Figure \ref{spectra}, the equivalent width (EW) of the He 1083~nm line appears to increase during the decay phase of the flare, whereas observations at other epochs exhibit stable EWs on similar timescales. We fit the EW increase in Figure~\ref{increase} with a linear model ($\mathrm{EW} = mt + b$) and find that $m = 0.66 \pm 0.19$ \AA$\:\mathrm{day^{-1}}$. Integrating the resulting posteriors for the slope $m$ suggests that $m>0$ at 99.9\% confidence. 

Because the flare occurred during the transit of V1298 Tau c, it is difficult to definitively say whether the increase in helium EW during the transit is due to the star or the planet. CARMENES observations of the helium line in M dwarfs have shown evidence for enhanced helium absorption during flares \citep{Fuhrmeister20}, consistent with our observations here. Recent work also suggests that stellar flares may lead to temporary enhancements in planetary mass loss rates and helium absorption; such enhancements are predicted to lag the peak of the flare by a few hours \citep[the dynamical timescale;][]{Wang21a}. Our observations were collected 0.5-1.5~hr from the flare peak, so we could also be observing enhanced helium absorption from planet c due to the flare. 

However, as we will discuss in Section~\ref{sec:bcdisc}, our non-detection of helium absorption in planet c allows us to place an upper limit on the magnitude of helium absorption from the planet's atmosphere. This may indicate that helium in this planet's atmosphere is already mostly ionized, in which case any additional high-energy input from the flare would be unlikely to enhance the level population in the metastable state. We conclude that the simplest explanation is an enhancement in chromospheric helium absorption, but this can be tested with additional helium transit observations of planet c.

% ------------------------------------------------------------------
\subsection{Non-Detections for V1298 Tau b and c}
\label{sec:bcdisc}
We detected the egress of V1298 Tau b with a modest transit-timing offset of $-57^{+14}_{-10}$~min. However, we did not require an extended planetary radius in the helium line to fit the data. Our upper limit of $\Delta R_\mathrm{b}/R_\star<0.019$ suggests that the planet does not exhibit strong helium absorption. This planet is relatively distant from its host star, so its atmosphere may not receive enough high-energy flux to drive a substantial outflow and/or create a substantial population of metastable helium. On the other hand, our data only cover half of the transit, which limits the sensitivity of our measurement. If planet b has an extended egress comparable to that of WASP-69 b \citep{Nortmann18} and WASP-107 b \citep{Spake21, Wang21b}, our data might still be consistent with a strong excess absorption signal during the transit. In this case the the transit-timing offset would need to be even larger, but we cannot exclude this possibility with the current partial phase coverage.

We were unable to detect the transit of V1298 Tau c at all. Our light curve exhibited strong correlated noise on the order of the broadband transit depth, likely stemming from the poor weather conditions during the observations. Our observations covered the full transit with approximately a transit duration's worth of additional baseline, so it would have required a TTV of $\gtrsim3~$hr to shift the transit out of the observation window. We conclude that the non-detection is likely due to correlated noise in our light curve. 

The non-detection of an escaping atmosphere is broadly consistent with recent work by \citet{Feinstein21}, who observed H$\alpha$ variations with Gemini-N/GRACES during a transit of V1298 Tau c but concluded that these variations were most likely stellar in nature as they did not appear to originate in the planetary rest frame. Our result suggests that either this planet has a relatively low mass loss rate, or the helium in the observable region of its atmosphere is mostly ionized. \citet{Oklopcic19} modeled planets orbiting older ($>$ Gyr) stars and found that planets on very close-in orbits should have helium ionization fronts at relatively low altitudes. Because ionization dominates over recombination through most of the upper atmospheres in these planets, the metastable state is not efficiently populated, and in-transit absorption in the 1083~nm line is suppressed. Although planet c's scaled semi-major axis is an order of magnitude larger than those of the planets with ionized atmospheres in \citet{Oklopcic19}, the star also outputs 100$\times$ more high-energy radiation \citep{Poppenhaeger21}, so strong ionization is a plausible explanation for the non-detection. A thin ionization front at low altitudes would also place this planet in the recombination-limited mass loss regime \citep{MurrayClay09, Owen16, Lampon21}. Consequently, the mass-loss efficiency (in the energy-limited formalism) should be quite low.

\subsection{A Tentative Signal for V1298 Tau d}
\label{sec:ddisc}
For planet d, we find stronger evidence for an extended atmosphere with $\Delta R_\mathrm{d}/R_\star = 0.018\pm0.005$ (3.6$\sigma$). However, the posterior requires a rather large transit-timing offset between the ingress and egress epochs of V1298 Tau d of $-100\pm20$~min. We note that the \textit{K2} data, which spanned a period of two~months, did not exhibit such large short-periodic TTVs. The data spanning ingress are well-behaved, with a scatter close to the photon noise limit and relatively little correlated noise. The transit time for this epoch was also independently measured with a broadband light curve (see \S\ref{sec:obs}), and the timing of ingress is tightly constrained by these data. On the other hand, the egress data have a larger variance relative to the photon noise limit and do not have any independent transit timing constraints.  It is therefore possible that our constraints on the egress time are somehow biased by correlated noise in our data; if this is the case, it would also affect our estimate of the transit depth in the helium band. We therefore consider the measured excess absorption signal to be tentative at best.

We tested the robustness of our fit by adding the PSF widths and centroid offsets as covariates with their own weights in our systematics model. The former covariate is particularly relevant for the night of the egress observation (November 27), which had relatively poor and variable seeing. We found that the best-fit weights for these parameters were indistinguishable from zero, yielding parameter estimates within 1$\sigma$ of those reported in Table~\ref{table1}. We conclude that our choice of systematics model has a negligible impact on the best-fit transit timing offset. 

We next consider whether or not telluric effects might cause a spurious signal in our data.  As discussed in \S\ref{sec:obs}, our helium bandpass overlaps with a strong telluric water feature, which can introduce a time-varying signal into our light curves \citep{Paragas21}. While we correct for this effect by using a proxy for the time-varying water column as a detrending vector for all observations, the correction may be inadequate if the planetary helium absorption signal overlaps with the telluric water lines and both are variable. For typical line widths, the water line will encroach on the planetary helium feature for geocentric velocities of 50~km/s $\lesssim v\lesssim$ 85~km/s. The geocentric velocity of V1298 Tau on our WIRC nights ranges between -8~km/s (for the 2020 October observations) and 33 km/s (for the 2021 January observations). However, for strong planetary outflows the line may be broadened \citep{Wang21b}, resulting in overlap with the telluric water features. Even at a modest 10~km/s geocentric velocity, the 2020 November HPF observations revealed that the strongly broadened stellar helium line was beginning to overlap with the telluric water lines. If the planetary helium line is similarly broad, then it is likely to overlap with the water line. Because the data for V1298 Tau d presented here are taken at small geocentric velocities, they should be less prone to this second-order effect. 

\subsection{Atmospheric Escape Modeling}
If we assume that the excess absorption signal observed for V1298 Tau d is due to the extended atmosphere of the planet, we can translate this quantity into an inferred mass loss rate. We compare our measured $\Delta R_\mathrm{d}/R_\star$ value to predictions from the atmospheric escape model described in \citet{Oklopcic18}. This model treats the outflow as a 1D Parker wind with a fixed mass-loss rate $\dot{M}$ and thermosphere temperature $T_0$, and compute the level populations and resulting helium absorption signal during transit. We explored models spanning a wide range of thermosphere temperatures
and mass loss rates, motivated by the broad range of values obtained in numerical simulations of atmospheric escape for a diverse sample of exoplanets presented in \citet{Salz16}. We mapped out the regions where the model predictions were consistent with our observations. For the photoionization and level population calculations we constructed an input stellar spectrum with an integrated XUV flux consistent with the estimate for V1298 Tau derived by \citet{Poppenhaeger21}, and a mid-UV flux consistent with spectra of T Tauri stars from \citet{Ingleby13}.  We tested our sensitivity to the assumed high-energy stellar spectrum by re-running the intermediate mass model grid with a different input spectrum \citep[also satisfying the integrated XUV flux estimate from][but with different wavelength dependence]{Poppenhaeger21}, and found that it shifted our inferred mass loss rates upward by approximately a factor of three; we conclude that the stellar spectrum is a comparable or smaller source of uncertainty than the assumed planet mass (see discussion below).  

\begin{figure}[t!]
\centering
\includegraphics[width=0.45\textwidth]{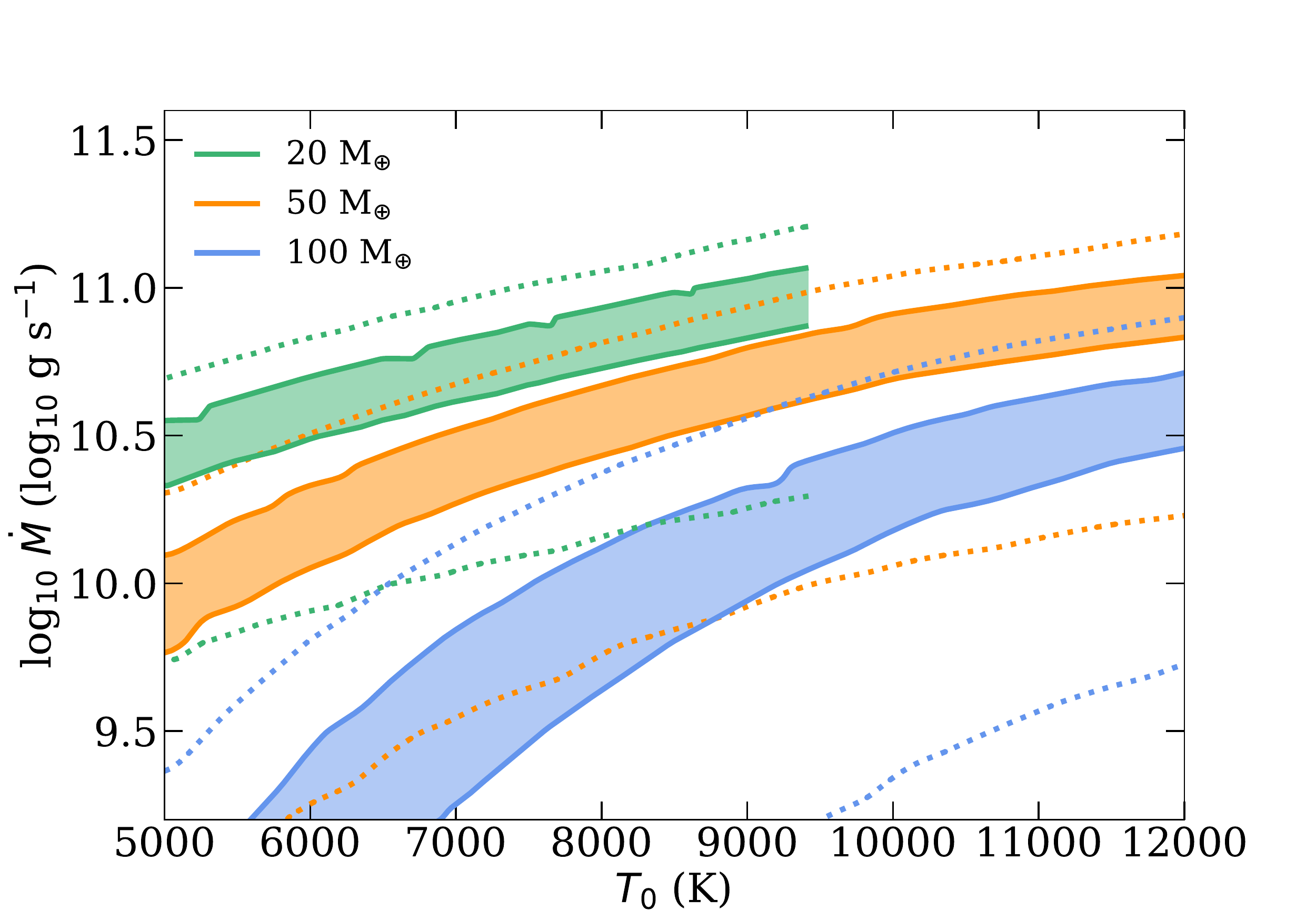}
\caption{Combinations of mass-loss rate $\dot{M}$ and outflow temperature $T_0$ that generate helium absorption consistent with our V1298 Tau d light curve. Solutions are calculated assuming the planet is $20M_\Earth$ (green), $50M_\Earth$ (orange), and $100M_\Earth$ (blue). Shaded regions denote 1$\sigma$ agreement with the observations and dotted lines denote 3$\sigma$ agreement.}
\label{masslossmodeling}
\end{figure}

In order to run these models, we must assume a value for the planet mass.  Unfortunately, there are no published measurements of the mass of V1298 Tau d. We therefore ran three separate model grids for representative masses of 20 $M_\Earth$, 50 $M_\Earth$, and 100 $M_\Earth$. The latter mass is a rough empirical upper limit from \citet{Thorngren19}, who studied an ensemble of older gas giant planets cooler than 1000~K. Planets in their sample with radii comparable to V1298 Tau d have $M \lesssim 100 M_\Earth$, so we take this as a conservative upper limit on the mass of planet d. On the low-mass end, previous studies of the V1298 system \citep{David19a, David19b} have suggested that these planets may have masses closer to those of the sub-Neptune population observed by Kepler (i.e., 1-10 $M_\Earth$).  This idea is supported by the current ensemble of TTV measurements for this system, which appear to favor $M_\mathrm{d} < 10 M_\Earth$ (Livingston et al. in prep.). For masses below 10 $M_\Earth$, the radius at which isothermal Parker wind becomes supersonic (assuming our standard range of thermospheric temperatures and the high ionization state of the atmosphere caused by the strong XUV flux of the host star) falls below the nominal radius of the planet. Since we do not consider our model assumption to be applicable in the low-mass part of the parameter space, we restrict our analysis to planet masses above 20 $M_\Earth$. We therefore use a 20 $M_\Earth$ model as our lowest mass case, and present the results in Figure~\ref{masslossmodeling}. 

Perhaps unsurprisingly, we find that there are a wide range of mass-loss rates that are consistent with our observations. For the nominal input spectrum, we can place a lower limit of $M/\dot{M} \approx 24$~Gyr (3$\sigma$) on the atmospheric lifetime for the hottest 20 $M_\Earth$ model. At a fixed thermosphere temperature, as the assumed mass increases the maximum mass-loss rate consistent with the data decreases, leading to even longer atmospheric lifetimes. This suggests that the atmosphere of V1298 Tau d should be stable against catastrophic mass loss if the planetary mass is indeed 20 $M_\Earth$ or larger. We can also estimate the threshold for catastrophic mass loss, here defined as $M/\dot{M} \lesssim 1$~Gyr, by assuming the largest mass-loss rate consistent with our observations still holds for $M_\mathrm{d} < 20 M_\Earth$. This approach suggests that the envelope can be fully removed if $M_\mathrm{d} \lesssim (10^{11.2}~\mathrm{g/s})(1~\mathrm{Gyr}) \approx M_\Earth$.  In reality, the crossover to catastrophic escape will happen at a larger mass as the mass-loss rate should be even larger for $M_\mathrm{d} < 20 M_\Earth$, but because we cannot model these cases with the 1D Parker wind methodology, we default to this conservative approximation.

\section{Conclusions} \label{sec:conc}
In this work, we searched for metastable helium in the atmospheres of V1298 Tau b, c, and d. We first characterized the stellar helium line using high-resolution spectra from HET/HPF.  We found that the helium line is relatively stable on hourly timescales in quiescent conditions, but can be highly variable timescales of days to months. We observed an appreciable increase in the helium equivalent width of spectra gathered during a transit of planet c, in the decay phase of a flare that was simultaneously observed photometrically with APO/ARCTIC. We concluded that this increase was most likely due to an increased population of metastable helium in the stellar chromosphere. We used diffuser-assisted narrowband photometry to measure light curves of V1298 Tau in a bandbass centered on the 1083~nm helium feature on four nights. We did not detect the transit of V1298 Tau c, and we modeled the transit of V1298 Tau b without needing an extended atmosphere. We found tentative evidence for planetary helium absorption in V1298 Tau d (3.6$\sigma$ significance), but the best-fit model required a relatively large transit-timing offset between the two transit epochs.

V1298 Tau is the only known young transiting system near a resonant chain, making it an important target for young planet studies. Ongoing TTV studies should eventually allow us to obtain well-constrained mass measurements (Livingston et al. in prep), therefore circumventing the difficulties of obtaining precise radial velocity measurements for this young, active star. TESS will observe V1298 Tau from 2021 September to 2021 November, significantly expanding the TTV baseline and providing improved dynamical constraints on the planet masses. These new observations will also help to reduce the growing uncertainty in the planetary ephemerides, which will be crucial for scheduling future ground-based transit observations. We expect that in the coming years, additional helium transit observations will be able to clarify the planetary origin of the tentative signal we report here. Ideally these observations should be obtained with long baselines and at geocentric velocities where the stellar helium line and the telluric water features do not overlap.  High resolution coverage of the metastable helium feature would also make it easier to disentangle the time-varying planetary, stellar, and telluric signals.

If the excess absorption signal from V1298 d can be confirmed at high signal-to-noise and combined with a well-constrained TTV mass measurement, our estimate of the planet's absolute mass loss rate will be limited by our knowledge of the star's high energy spectrum. \citet{Poppenhaeger21} measured V1298 Tau's X-ray spectrum between 0.1 keV and 2 keV, but the photoionization physics for metastable helium is governed primarily by mid-UV and EUV flux near the ionization thresholds of the metastable and ground states, respectively \citep{Oklopcic18, Oklopcic19}. One way of reconstructing the EUV spectrum is the differential emission measure technique, which relies on measurements of stellar emission lines in the FUV \citep{Duvvuri21}. Although it would be challenging to detect these lines in V1298 Tau's spectrum, as this star is located at a distance of 108.5~pc, they may be observable with the \textit{Hubble Space Telescope}.

Constraining mass-loss rates in a young, well-characterized multi-planet system remains an important goal. These measurements can help differentiate between the recombination-limited, energy-limited, and photon-limited regimes for mass loss at early times, which would have crucial implications for the outflow efficiencies \citep{Lampon21}. In turn, this would allow us to benchmark population-level mass-loss models, like those used to infer core masses and compositions of the \textit{Kepler} planets \citep{Rogers21}. As the precision of mass-loss measurements improve for younger planets, V1298 Tau will be a keystone system for understanding and characterizing this crucial process when it is most important for planetary evolution.

\acknowledgments
We thank the Palomar Observatory staff and directorate for their herculean effort to establish safe, remote operations during the COVID-19 pandemic. In particular we thank Kajse Peffer, Carolyn Heffner, Joel Pearman, Paul Nied, Kevin Rykoski, and Tom Barlow for telescope operations and remote support, and we thank Andy Boden for facilitating scheduling of these observations. SV is supported by an NSF Graduate Research Fellowship and the Paul \& Daisy Soros Fellowship for New Americans. HAK acknowledges support from NSF CAREER grant 1555095.

This work was partially supported by funding from the Center for Exoplanets and Habitable Worlds. The Center for Exoplanets and Habitable Worlds is supported by the Pennsylvania State University, the Eberly College of Science, and the Pennsylvania Space Grant Consortium. This work was supported by NASA Headquarters under the NASA Earth and Space Science Fellowship Program through grants 80NSSC18K1114. We acknowledge support from NSF grants AST-1006676, AST-1126413, AST-1310885, AST-1517592, AST-1310875, AST-1910954, AST-1907622, AST-1909506, the NASA Astrobiology Institute (NAI; NNA09DA76A), and PSARC in our pursuit of precision radial velocities in the NIR. We acknowledge support from the Heising-Simons Foundation via grant 2017-0494 and 2019-1177. Computations for this research were performed on the Pennsylvania State University’s Institute for Computational \& Data Sciences (ICDS).

These results are based on observations obtained with the Habitable-zone Planet Finder Spectrograph on the Hobby-Eberly Telescope. We thank the Resident astronomers and Telescope Operators at the HET for the skillful execution of our observations of our observations with HPF. The Hobby-Eberly Telescope is a joint project of the University of Texas at Austin, the Pennsylvania State University, Ludwig-Maximilians-Universität München, and Georg-August Universität Gottingen. The HET is named in honor of its principal benefactors, William P. Hobby and Robert E. Eberly. The HET collaboration acknowledges the support and resources from the Texas Advanced Computing Center. 

\facilities{Hale (WIRC), HET (HPF), ARC 3.5m (ARCTIC), ADS, Exoplanet Archive}
\software{\texttt{allesfitter} \citep{gunther2019, gunther2021}, \texttt{astropy} \citep{Astropy13, Astropy18}, \texttt{celerite} \citep{foremanmackey2017}, \texttt{celerite2} \citep{ForemanMackey18}, \texttt{exoplanet} \citep{ForemanMackey20}, \texttt{lightkurve} \citep{lightkurve}, \texttt{matplotlib} \citep{Hunter07}, \texttt{molecfit} \citep{smette2015,kausch2015}, \texttt{numpy} \citep{Harris20}, \texttt{photutils} \citep{Bradley19}, \texttt{pymc3} \citep{Salvatier16}, \texttt{scipy} \citep{Virtanen20}, \texttt{starry} \cite{Luger19}, \texttt{theano} \citep{theano16}}

\allauthors 

\end{document}